\documentclass[twocolumn]{revtex4}

\usepackage{graphics}
\usepackage{graphicx}
\usepackage{dcolumn}
\usepackage{bm}
\usepackage{epsfig} 
\usepackage{psfrag} 

\usepackage[dvips]{color}

\newcommand{\beq}{\begin{equation}}
\newcommand{\eeq}{\end{equation}}
\newcommand{\D}{| \psi |}
\newcommand{\Db}{| \bar{\psi} |}

\newcommand{\R}{| \tilde \psi |}

\newcommand{\lt}{\left}
\newcommand{\rt}{\right}

\newcommand{\smallequa}{\ \lower-1.2pt\vbox{\hbox{\rlap{$<$}\lower5pt\vbox{\hbox{$\sim$}}}}\ }

\begin{document}


\title{Pseudogap and Amplitude Fluctuations in High Temperature Superconductors}

\author{Philippe Curty$^1$ and Hans Beck$^2$ }
\affiliation{ $^1$Scimetrica Research, 3007 Bern, Switzerland\\ $^2$ Universit\'e de Neuch\^atel, 2000 Neuch\^atel, Switzerland}
\date{\today}
\begin{abstract}
Amplitude fluctuations of the pairing field are responsible together with phase fluctuations for the pseudogap phenomena in high temperature superconductors.
 Here we present the more detailed theory of the amplitude and phase fluctuations approach in the framework of a fermionic pairing model.
 New experimental comparisons are presented for the specific heat of the curprate LSCO confirming the generality of this phenomenological
 approach. The strong decrease of amplitude fluctuations near optimal doping induces the illusion of a "quantum critical point", which in fact does
 not exist since the pseudogap energy scale is always different from zero even in the overdoped regime.
\end{abstract}
\maketitle

One major problem concerning high temperature superconductors is to establish the correct phase diagram temperature versus hole or electron
 doping.  A related question is the interpretation of the anomalous behaviour above the critical temperature. This regime is called pseudogap
  \cite{loram94,timusk} because it contains effects similar to superconductivity like a partial suppression of electronic density of states.
The pseudogap region starts below a temperature $T^*$ where observable quantities deviate from Fermi liquid behaviour and seems to
be present well inside the superconducting phase.

   Until now two interpretations seem to emerge: the first point of view assumes the existence of quantum critical point (QCP) sitting
   in the middle of the superconducting regime \cite{sachdev}. This quantum critical point is due to an (unknown) order parameter which
   is competition with superconductivity. The experimental evidences for a QCP
   are however not convincing until now. Moreover  effects of quantum fluctuatations are suppressed by superconductivity at low temperature.
   Loram and Tallon \cite{loram2} have analyzed many experimental data like specific heat or scanning tunneling experiments and extract the
   value of an external energy scale $E_g$. In their analysis, they assume the presence of an external gap and extract its value by fitting
   experiments.

The second group of approaches tends to analyse the consequence of pairing and superconductivity on the phase diagram
\cite{emery,xu2000,curty2003}. The pseudogap phase is then due to precursor effects of  superconductivity or amplitude pairing
fluctuations for example. With this point of view, one does not need the existence of the QCP. However, there is still the need to
explain the characteristic pseudogap energy found  $E_g$ by Loram and Tallon \cite{loram2}.

It is perhaps important to remind that $T^*$ is a crossover temperature that can be directly seen in experiments where observable
 quantities deviate from Fermi liquid behaviour . $E_g$ in the sense of ref. \cite{loram2} is a phenomenological energy
  scale extracted by fitting experiments according to a simple theory of gapped electrons.

\begin{figure}[ht]
\centering
\resizebox{8.5cm}{!}{\includegraphics{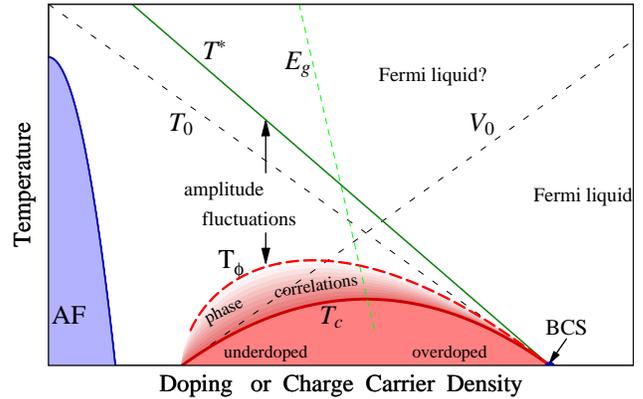}}
\caption{\label{PhaseDiagramCuprates} Schematic phase diagram of cuprates. The pseudogap region of the copper oxides phase diagram
 lies between the critical temperature $T_c$ and a temperature $T^*$.  The temperature $T_{\phi}$ has been reported in several experiments  \cite{wang,naqib} and shows the onset of phase correlations or superconducting fluctuations. The energy scale $E_g$ has also been found in
   many experiments like specific heat  \cite{loram2} or transport \protect{\cite{naqib}}. $V_0$ and $T_0$ are the model parameters. $V_0$ is the superfluid stiffness and $T_0$ the mean-field pairing temperature.}
\end{figure}
In a recent letter \cite{curty2003}, the importance of treating both amplitude and phase fluctuations in a pairing model has been shown.
 Using this method, it is possible to fit accurately experiments and to extract all the phase diagram of cuprates except the antiferromagnetic
  regime. By expanding observables around the average amplitude, one can compute thermodynamic quantities like specific heat or magnetic
   spin susceptibility.  Two  regions have to be distinguished (see Fig. \ref{PhaseDiagramCuprates}): for a
relatively small temperature interval $T_c < T < T_{\phi}$ the phase of  $\psi$ is still correlated in space over some coherence length $\xi$
 (the Kosterlitz-Thouless coherence length in $2 d$).
 Thus, in this regime, observables are governed by correlated phase fluctuations described by the XY-model. For $T_{\phi} < T <
T^*$, phases of $\psi$ are essentially uncorrelated ($\xi$ is on the order of the lattice constant), but $\D$ is still fluctuating and non-zero,
signaling local pair fluctuations. This explains the wide hump between $T_c$ and temperature $T^*$ seen in specific heat experiments \cite{loram94}, the
depression of the spin susceptibility \cite{takigawa} and the persistence of the pseudogap for $T < T^*$.

 One important result is the extracted phase diagram: by fitting the experiments it is possible to get the phase stiffness $V_0$,
  the mean-field temperature $T_0$ and the energy scale $E_g$ versus doping.\\

In this paper, we will show that the phase diagram of cuprates has indeed an energy scale $E_g$ simililar to the one found in  \cite{loram2}.
However this pseudogap energy scale is due the presence of a pairing attraction and controlled by amplitude fluctuations.
When these two effects are taken into account, they produced a very large $E_g$ in the underdoped regime where the superfluid
 density is low and fluctuations are large. In the maximum $T_c$ regime, $E_g$ becomes rapidely small because amplitude fluctuations
  decrease rapidely. The interaction strenth is also decreasing when overdoping. This creates the illusion of a QCP in the middle of the superconducting region.

Our approach has the advantage that the only strong assumption is that high temperature superconductivity is caused by a fermionic
 pairing attraction which seems to be likely the case. All other assumptions are only technical and related to the calculations.

Various experimental observations can indeed be interpreted in terms of fluctuations of the pairing field $\psi = |\psi| e^{i \phi}$, and that
two temperature regions have to be distinguished (see Fig. \ref{phasediagram}): for a
relatively small temperature interval $T_c < T < T_{\phi}$ the phase of  $\psi$ is still
correlated in space over
some coherence length $\xi$ (the Kosterlitz-Thouless coherence length in $2 d$)
whereas the amplitude $\D$ is almost constant. Thus, in this regime, observables are
governed by correlated phase fluctuations described by the XY-model. For $T_{\phi} < T <
T^*$, phases of $\psi$ are essentially uncorrelated ($\xi$ is on the order of the
lattice constant), but $\D$ is still non-zero, signaling independent fluctuating local
pairs. This explains the wide hump seen in specific heat experiments \cite{loram94}, the
depression of the spin susceptibility \cite{takigawa} and the persistence of the pseudogap for $T < T^*$.
Moreover, a magnetic field that destroys this pseudogap
has to break fluctuating pairs and must therefore be much higher than the one which
suppresses phase coherence and thus superconductivity \cite{pieri}.

Our approach has a major difference with the Emery and Kivelson phase fluctuations scenario \cite{emery} of the pseudogap regime:
our calculations show that phase fluctuations influence the pseudogap only up to a temperature  $T_{\phi} $ which is much smaller
 than $T^*$. Above $T_{\phi}$, observables are thus only determined by the amplitude of the pairing field.

\index{amplitude fluctuations}
\begin{figure}[floatfix]
\includegraphics[width=0.6\columnwidth]{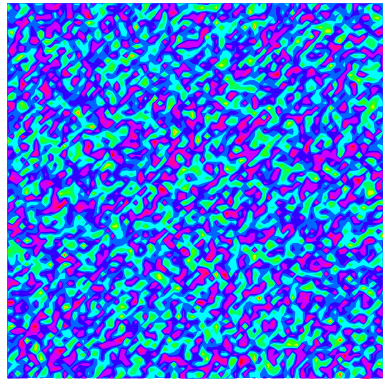}
\caption{Amplitude of the pairing field  above $T_c$ on a $80\times80$ array by cluster Monte Carlo
 simulations: dark color represents large amplitude and light colors low amplitude.}
\label{distribution}
\end{figure}

The existence of a temperature intermediate between $T_c$ and $T^*$ has also been
mentioned by Devillard and Ranninger \cite{devillard}: using a Boson-Fermion
 description of pairing in cuprates, they find that uncorrelated pairing of
electrons leads to the opening of a pseudogap at $T^*$.
 These pairs acquire well behaved itinerant features at $T_B^*$,
leading to partial Meissner screening, and thus to diamagnetic susceptibility, and
Drude-type behaviour of the optical conductivity. As a function of lattice anisotropy (and thus of doping) $T_B^*$
has the same tendency as $T_c$, whereas the higher temperature $T^*$ has the opposite trend. Although in ref
\cite{devillard} $T_B^*$ is related to the temperature where the pair life time becomes long, it could be identified with our $T_{\phi}$.

\section{Model and Effective Action}
We are not primarily interested in the origin of pairing between electrons but in the
consenquences of pairing on the phase diagram or physical quantities like specific heat or magnetic susceptibility.
 We assume that there is an  attractive force between fermions,
and we base our calculations on a $d$-wave attractive Hubbard model
\begin{equation}
\label{Hamiltonian}
H = -\sum_{\lt<i,j \rt> \sigma} t \ c^{\dagger}_{i \sigma} c_{j \sigma}
- U \sum_{i} Q_{d}^{\dagger}(i) \ Q_{d}(i)
\end{equation}
with a hopping $t$ between nearest neighbour sites $i$ and $j$ on a square lattice.
The interaction favours the formation of onsite $d$-wave pairs since
\beq
Q_{d}^{\dagger}(i) =  \sum_j D_{ij} \ Q^{\dagger}_{ij}
\eeq
where $D_{ij}=1, (-1)$ for $i$ being the nearest neighbour site of $j$ in horizontal
(vertical) direction.
\beq
Q_{ij}^{\dagger} = \lt( c_{i \uparrow}^{\dagger} c_{j
 \downarrow}^{\dagger} - c_{i \downarrow}^{\dagger} c_{j \uparrow}^{\dagger}\rt)/\sqrt{2}
\eeq
is an operator creating a singlet pair on neighbouring sites.
Decoupling the interaction with the help of a Stratonovich-Hubbard
transformation, the partition function $Z = \mbox{Tr} \ e^{- \beta H}$ is then
\begin{equation}
\nonumber
 Z =  Z_n \int D^2 \! \psi \lt< {\cal T} e^{-  \int_0^{\beta} d\tau
\sum_{i} \lt(\frac{1}{U} |\psi|^2 +  \psi \ Q_d^{\dagger}(i) + \mbox{\small
hc} \rt)} \rt>_{H_n}
\end{equation}
where 
\beq
\psi = \psi(i, \tau)= |\psi(i,\tau)| e^{i \phi(i, \tau)},
\eeq
 and $H_n=-\sum_{\lt<i,j \rt> \sigma} t \ c^{\dagger}_{i \sigma} c_{j \sigma}$ is the
non-interacting part. The trace over the fermionic operators can be evaluated yielding
\begin{equation}
\label{Z.fermionic}
Z = \int D^2 \! \psi \ e^{-\int_{0}^{\beta}  d \tau \lt[ \sum_i \frac{1}{U}
|\psi|^2  + \mbox{Tr} \ln G \rt]}.
\end{equation}
Here $G$ is a Nambu matrix of one-electron Green functions for fermions interacting with
a given, space and time
dependent pairing field $\psi(i, \tau) $. The Green functions are solution of Gorkov's equations (see \cite{gyorffy}).

Expanding (\ref{Z.fermionic}) in power of $\vec{\nabla} \psi$, $Z$ can be written as a
functional integral involving an action $S[\psi]$ for a field $\psi$ that changes slowly in
space and that can be taken time-independent: \index{effective action}
\beq
S[\psi] =  S_0(\D) + S_1( \vec{\nabla}\psi )
\label{action}
\eeq
where $S_0$ is a local functional of $\psi$:
\beq
S_0(\D) = V \ {\D^2 \over U}  - {2 \over \beta } \sum_q  \log
 [ 2 \cosh { \beta E_q/2} ]
 \eeq
and $S_1 = c \int d^3 \! r {|\vec{\nabla} \psi |^2 / 2}$ can be considered as the deformation or kinetic energy where $c$ is a constant.
The quasi-particle energy is $E_q = \sqrt{\D^2+(\varepsilon_q -\mu)^2}$.\\

\index{thermodynamics}
Now we would like to compute thermodynamic observables such as energy $ U = \langle  S \rangle_{S}$,
 specific heat and spin susceptibility. The point is that we want to keep the XY universality class of the transition together with the
 fermionic character of the system: in the limit of high density or weak interaction, the superconductor should be described by a
 BCS like mean field theory whereas in the low density limit with strong interaction the transition becomes XY like. Our goal is
 to derive a theory that describes these two regimes and, of course, the intermediate regime.

Our  main strategy will be to neglect amplitude
correlations since simulations show that they are weak between different sites $i, j$: $\langle \D_i \D_j \rangle - \langle
\D^2 \rangle \approx 0$ since the amplitude is always positive and cannot show any critical
behaviour. 
In this spirit, two different approaches are possible:\\

First approach:  the amplitude is fixed  but still temperature dependent, and is determined by a suitable variational equation. Fluctuations
from the XY model are kept as well as the amplitude weight coming from the Jacobian of the cartesian to polar coordinate transformation.\\

Second approach:  the energy is expanded around the average amplitude. Higher powers of amplitude fluctuations are neglected. Here the
local coupling between phase and amplitude is kept, and amplitudes are allowed to fluctuate.\\

\section{Variational Method} \index{variational method}
The integration of the partition function can be expressed in polar coordinates using the transformation:
\beq
D\psi = \prod_i  \int_{-\infty}^{+\infty} d\psi_i \ d\psi^*_i  = \prod_i  \int_{0}^{+\infty} d|\psi_i| |\psi_i| \  \int_{0}^{2 \pi} d\phi_i
\eeq
Rewriting the free energy $F$ in terms of a constant amplitude $\D$ yields
\beq
\nonumber
F = -{1 \over \beta} \log  \int D \phi  \ e^{-\beta \lt( S_0(\D) - \log(\D)
V/\beta +
 S_1\rt) }
\eeq
where the Jacobian $|\psi_i|$ of the polar transformation is put into the exponential, and $V$ is the
volume. Taking the derivative of $F$ with respect to $\D$ and equating it to zero leads to
the self-consistent equation:
\beq
{\partial S_0(\D) \over \partial \D } - {V \over \beta \D} + c  \D \ \lt< | \vec{\nabla}
e^{i \phi}
 |^2
\rt>_{S_1} = 0.
\label{amplitudeequation1}
\eeq
Evaluating equation (\ref{amplitudeequation1}) and multiplying it with  $\D/2$
\begin{eqnarray}
{\D^2\over U} -   { \D^2 \over W}  \int_{-\mu}^{W-\mu}    \! d\xi \
{
\tanh \lt(\sqrt{\xi^2+ |\psi|^2}/(2T)\rt)
\over
\sqrt{\xi^2+ |\psi|^2} \nonumber
}\\
 - {1 \over 2 \beta }
+ c \D^2  \ \lt< | \vec{\nabla} e^{i \phi}|^2 \rt>_{S_1} = 0.
\end{eqnarray}
\label{amplitudeequation2}
The first and second terms are the amplitude contribution and leads to the BCS gap equation if other contributions are neglected.
The third comes from the Jacobian and  implies that the amplitude is never zero. The last term is the expectation
value of the energy $U_{xy}$ in the XY  model with a constant dimensionless coupling $K$
\beq
 K = {V_0 \over T} {\D^2 \over |\psi_0|^2} .
\label{couplingK}
\eeq
where $V_0$ is the zero temperature phase stiffness, \index{phase stiffness} and $|\psi_0|$ is the zero temperature amplitude .
This contribution characterises the influence of the phase fluctuations.  $U_{xy}(K)$ is a monotonic decreasing function with an inflexion point at $T_c$.
Solutions of equation (\ref{amplitudeequation2}) are reliable for all temperatures except for
$T<<T_c$. However they are only expected to be accurate at $T_c$ if the average amplitude is large and not varying to much with temperature.
\begin{figure}
 \setlength\unitlength{1cm}
\begin{center}
 \begin{picture}(8.5,5.5)
\put(0.25,0){\resizebox{8cm}{!}{\includegraphics{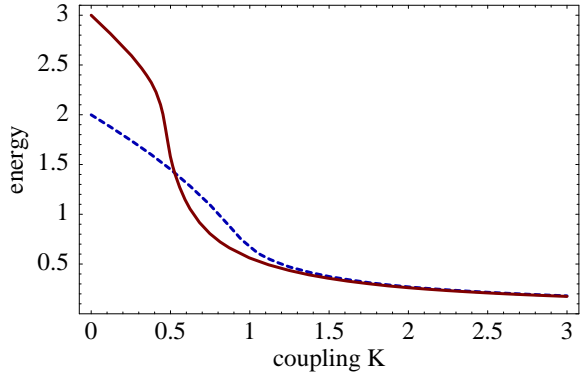}}}
\end{picture}
\end{center}
\caption{Dimensionless energy of the XY model as a function of the coupling $K$ defined in
equation (\protect{\ref{couplingK}}). The continuous line is for 3 dimensions and dashed is
for 2 dimensions. Note that the phase transition occurs near the cusps.}
\label{Uxy}
\end{figure}

\noindent
\section{Average Value Method} \index{average value method}
Since simulations show that amplitude correlations are weak, we can expand the energy around the average amplitude
where it is not coupled to the phase. The energy is then
\beq
U = \langle  S \rangle_{S} \approx S_0(\langle \D \rangle) + \langle  S_1 \rangle_{S}
\eeq
Here, $ S_0(\langle \D \rangle) $ is the first term of an expansion of the average $\langle S_0(\D) \rangle$
around $\langle \D \rangle$.
\begin{eqnarray}
\langle  S \rangle_{S} =&&
\lt<
S_0(\langle \D \rangle) + c_1 \delta|\psi| + c_2 {\delta|\psi|^2 \over 2} + ...
\rt> \nonumber \\
=&&  S_0(\langle \D \rangle) +
 \sum_{n=1}^{\infty} { c_n  \over n!} { \langle \delta|\psi|^{2n} \rangle }
 \label{averageS}
\end{eqnarray}
where $\delta|\psi| =\D -\langle \D \rangle$.
Assuming that amplitudes have gaussian fluctuations around their average value, we can use the Wick's theorem:
 $$\langle (\D -\langle \D \rangle)^{2n} \rangle = (2n+1)!! (\langle \D^2 \rangle -\langle \D \rangle^2)^{n}$$
This last identity together with equation (\ref{averageS})  leads to
\begin{eqnarray}
\langle  S \rangle_{S} = \sum_{n=1}^{\infty} c_n { (2n+1)!! \over n!}  (\langle \D^2 \rangle -\langle \D \rangle^2)^{n}
\end{eqnarray}
Higher corrections term are proportionnal to powers of the square of the standard deviation
$\langle \lt| \psi \rt|^2 \rangle - \langle \lt| \psi \rt| \rangle^2$. Although amplitude fluctuations are large,
the standard deviation is half of the average amplitude, powers of  $\langle \lt| \psi \rt|^2 \rangle - \langle \lt| \psi \rt| \rangle^2$
are small compared to the average amplitude itself. This is why we only take the first term of the
expansion which already contains the main thermodynamic features.
Additional terms would just add more fluctuations and small corrections to  the results.
\begin{figure}
 \setlength\unitlength{1cm}
\begin{center}
 \begin{picture}(8,5.5)
\put(0.5,0){\resizebox{7cm}{!}{\includegraphics{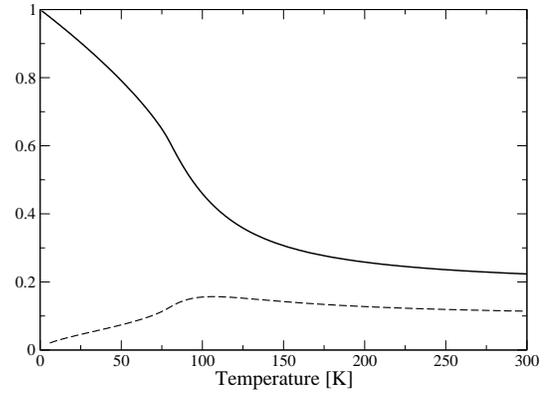}}}
\end{picture}
\end{center}
\vspace{0cm}
\caption{Average amplitude (thick line) and the corresponding standard deviation (dashed line). Parameters are
$V_0 = 215$, $T_0 =140$. Note that $T_c$ is about 80K whereas the amplitude remains non zero at least up to 300K.}
\label{fig-amplitudedeviation}
\end{figure}
\index{amplitude fluctuations}
\begin{figure}
 \setlength\unitlength{1cm}
\begin{center}
 \begin{picture}(8,5.5)
\put(0.5,0){\resizebox{7cm}{!}{\includegraphics{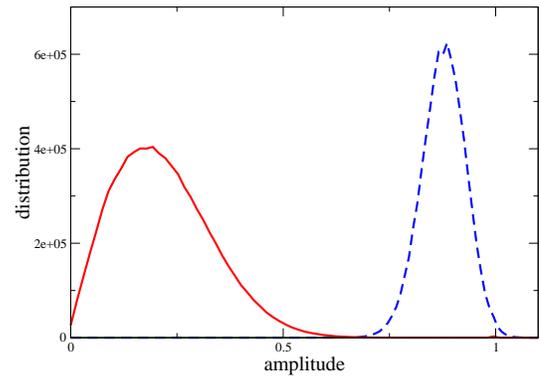}}}
\end{picture}
\end{center}
\vspace{0cm}
\caption{Amplitude distribution for $T=300$K (thick line) and $T=30$K (dashed line). Parameters are
$V_0 = 215$K, $T_0 =140$K. Note the broadening of the amplitude distribution at high temperature, i.e the
increase of fluctuations, compared to the low temperature distribution where the distribution is narrower.}
\label{fig-amplitudedistribution}
\end{figure}

For simplicity, averages are computed using a normalised Ginzburg-Landau action $S_{GL}$
(see \cite{curty}) 
\index{Ginzburg-Landau}
whose potential part $U_{GL}$ is equal to the first two terms of the expansion of $S_0$
with respect
to $\beta \D $:
\beq
 S_{GL}[\psi] = k_B V_0 \int{d^3 \! r  \lt(  U_{GL}  +  S_1 \rt)}
\label{HGL2}
\eeq
where 
\beq
U_{GL} = \eta^2 ( t  |\tilde \psi| ^2+ {1\over 2} { |\tilde \psi|^4} ),
\label{GL_potential}
\eeq
and $t \approx T/T_0-1$ is the reduced temperature, $ \tilde \psi = {\psi / | \psi(T=0) |}$ is the reduced field, $T_0$
is the mean field BCS  pairing temperature. $S_{GL}$ is normalised with a lattice spacing  $\varepsilon$.
$$ 
\eta := \varepsilon / \xi_0,
$$
where $\xi_0$ is the mean field coherence length at zero
temperature, and  $V_0$ is the zero temperature phase stiffness.
Contrary to the variational method, amplitude and phase are still {\em locally} coupled 
through $S_1$. The energy becomes
\beq
U \approx S_0(\langle \D \rangle_{GL}) + \langle S_1 \rangle_{GL}
\label{energy}
\eeq
where
$$
S_0(\langle \D \rangle) = ( V \ {\langle \D \rangle^2 / U})  - {2 \over \beta }\sum_q  \log
 [ 2 \cosh { \beta E_q/2} ]
$$
corresponds the BCS free energy for which the gap value is determined by
the GL average. The quasi-particle energy is $ E_q = [ (\varepsilon_{q}-\mu)^2 +
 \langle \D \rangle^2 \cos^2(2\theta) ]^{1/2} $ where $\mu$ is the chemical potential.
  The $d$-wave symmetry manifests itself by the angle dependent amplitude
  $\D \cos(2 \theta)$ where $\theta$ is the angle in $q$ space with respect to $q_x$
direction.

 The value of $\eta$ depends on the corse-graining procedure and is fixed for each sample.
Observables are not very sensitive to changes in $\eta$.

Both approaches are valid below and above  the critical temperature $T_c$ which is the
temperature where the phase stiffness becomes zero.
However the average value method gives good results for all values of $T_0$ and $V_0$ whereas the variational
method works better in the underdoped regime, i.e. for $V_0 < T_0$.

It is important to notice that the amplitude is fluctuating although it is fixed to its averaged value in observables.
In figure \ref{fig-amplitudedeviation}, the average amplitude and its standard deviation are shown. One can note that
the latter is approximately half of the averaged amplitude.  Since the amplitude distribution around its average value is
almost gaussian, the probability density $p(|\psi|)$ has the following form:
\beq
p(|\psi|) \approx e^{- ( |\psi| - \langle |\psi| \rangle)^2 \over \langle |\psi| \rangle }
\eeq
This means that the amplitude has values ranging from  0 to $2  \langle |\psi| \rangle$ causing the large average value $\langle |\psi| \rangle$.\\

Computer simulations of the statistical ensemble $\{\psi\}$ under the action
 $S_{GL}$ have been done using a standard Monte Carlo procedure to update amplitude $\R$ and a Wolff
\cite{wolff} algorithm for the phase $\phi$ in the same way as for the real $\Phi^4$ model
\cite{brower}. Typically $10^4$ sweeps are needed to obtain good statistics.
.

Since the fitting procedure is done completely automatically, one has to simulate first the Ginzburg-Landau action
 (\ref{HGL2}) for all possible parameters $V_0$ and $T_0$. Once a mesh of simulations has been done, it is possible
 to interpolate the surface defined by $\langle |\psi| \rangle(V_0,T_0)$ and to have access all values of $\langle |\psi| \rangle(V_0,T_0)$.
The same is done for the energy $\langle | \nabla \psi|^2 \rangle(V_0,T_0)$. This allows to have an efficient, automatic and reproducible
fitting procedure.

\begin{figure}[ht]
\begin{center}
 \setlength\unitlength{1cm}
 \begin{picture}(8.5,5)
 \put(0,-0.5){\resizebox{8.5cm}{!}{\includegraphics{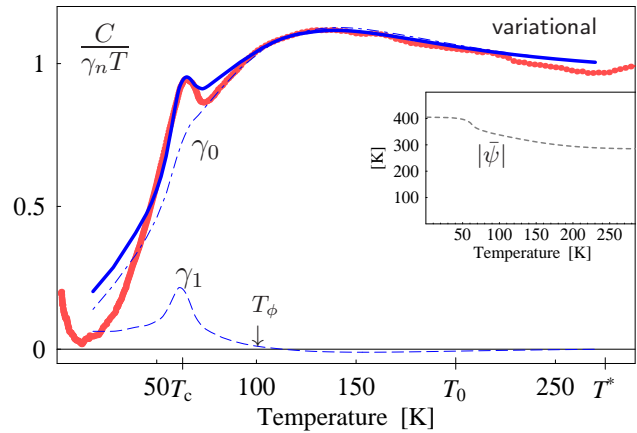}}}
 \put(6.4,4.8){ {\small \sf  variational} }
 \put(1,4.5){\Large $ C \over \gamma_n T$}
 \put(2.3,1.5){\large $ \gamma_1$}
 \put(3.2,1.1){ $T_{\phi}$}
 \put(3.2,0.7){ $\downarrow$}
 \put(2.4,3.2){ \large $\gamma_0$}
  \put(6.3,3.1){\small $\Db$}
 \end{picture}
\end{center}
\caption{ The reduced specific heat $\gamma$ from the variational method (thick),
 which is the sum of the gradient $\gamma_1$ (dashed) and amplitude $\gamma_0$ (dotted-dashed) contributions,
 reproduces measurements  of YBa$_2$Cu$_3$O$_{6.73}$ (points). {\em Inset:} The dashed line is the temperature
  dependent amplitude $\Db$ from equation (\protect{\ref{amplitudeequation2}}). }
\label{specificheat1}
\end{figure}
\noindent
\section{Specific Heat}
\index{specific heat}
In both approaches, the specific heat $C$ is  the sum of the amplitude $C_0$ and the gradient $C_1$ contributions.
 Defining the reduced specific heat  $\gamma = C/(\gamma_n T)$,  we have
\beq
\gamma = \gamma_0  + \gamma_1
\eeq
where $\gamma_1$ is divided by $T_c$ instead of $T$ since $S_1$ is classical
 and does not satisfy the third law of thermodynamics.

In practice, the amplitude specific heat can be calculated by using the entropy. $C_0$ is proportional to the derivative of
the entropy times the temperature:
\beq
C_0(T,|\psi|) = -T  {\partial S(|\psi|/T) \over \partial T}
\eeq
where
the amplitude $|\psi|$ is replaced by $|{\bar \psi}|$ in the variational approach and by $ \langle |\psi| \rangle$ in the average value approach.
The entropy $S$ for a fermionic system in the presence of a gap $|{ \psi}|$  is a universal function of the ratio $|{ \psi}| /T$:
\beq
S(|{ \psi}| /T)  =  -k_B \sum_{k}   f_k \log(f_k)
\eeq
where the Fermi distribution is $f_k= {1 \over e^{-\beta (E_k -\mu)} +1}$ and the energy $E_k$
is $E_k= \sqrt{\xi_k + |\psi|^2}$. By using the entropy, it is not necessary to perform the sum
over $k$ each time one wants to evaluate the specific heat. It is then sufficient to take the
derivative of the entropy.

{\bf Normalisation of the specific heat:}
$\gamma_0$  is 1 at high temperature since it is divided by the Sommerfeld constant $\gamma_n$.
The phase contribution is normalised as:
\beq
{C_1 \over \gamma_n} = {k_B \over \xi_0^3 \gamma_n} {C^{(s)}_{\phi} \over N k_B}
\eeq
where $C_1$ is the specfic heat per volume $V=N \xi_0^3$, and $C^{(s)}_{\phi}/(N k_B)$ is the specific heat
 per number of lattice sites  coming from the simulations. Experiments give $\gamma_n \approx$  26 mJ K$^{-1}$
mol$^{-1}$ = 252 J K$^{-1}$ m$^{-3}$. For the fit of Fig. \ref{specificheat1}, using the reasonable value $ \xi_0 \approx 16$ \AA,
 we get  the dimensionless constant $\alpha={k_B /( \xi_0^3 \gamma_n}) \approx 13.5$.

\subsection{Specific heat for the variational method (d-wave)}

\noindent
Using the variational method the specific heat $C$ is now
\beq
\gamma = \gamma_0\lt( \Db  \rt) + \gamma_1
\eeq
where the amplitude $\Db$ is now solution of the amplitude equation (\ref{amplitudeequation2})
The dimensionless constant is $\alpha \approx 13.5$.

In Fig. \ref{specificheat1} the experimental specific heat of YBa$_2$Cu$_3$O$_{6.73}$ \cite{loram94} is fitted using the
 variational method reproducing the double peak structure: a sharp peak below $T_{\phi}$ coming from phase fluctuations
  and a wide hump below $T^*$ rounded by amplitude fluctuations. The crossover temperature $T_{\phi}$,  where phases
   become random, corresponds to the temperature where $\gamma_1$ is less than approximately 2\% of the  normal
   specific heat. In  amplitude equation (\ref{amplitudeequation2}), a 2 dimensional density of states
$D(\varepsilon) = 1/W$ is used with $W=5000 \mbox{K}$, $\mu = 0.25 W$ and  $U=959 \mbox{K}$. These parameters
 gives $T_0 \approx 200 \mbox{K}$ and $\psi_0 \approx 2.14 T_0$ in agreement with experiments \cite{kugler}.
  The other parameters are $V_0 = 72 \mbox{K}$ and $\eta=5$.

\subsection{Specific heat for the average value method ($d$-wave)}
Following equation (\ref{energy}), the specific heat is given by
\beq
\gamma = \gamma_0\lt(\langle \D \rangle_{GL} \rt) + \gamma_{1}
\eeq
where $\gamma_{1}$ is divided by $T_c$ instead of $T$ since $S_{GL}$ is a classical action and does not satisfy the Nernst theorem.

\begin{figure}[ht]
\begin{center}
 \setlength\unitlength{1cm}
 \begin{picture}(8.5,5.5)
 \put(0,-0.5){\resizebox{8.5cm}{!}{\includegraphics{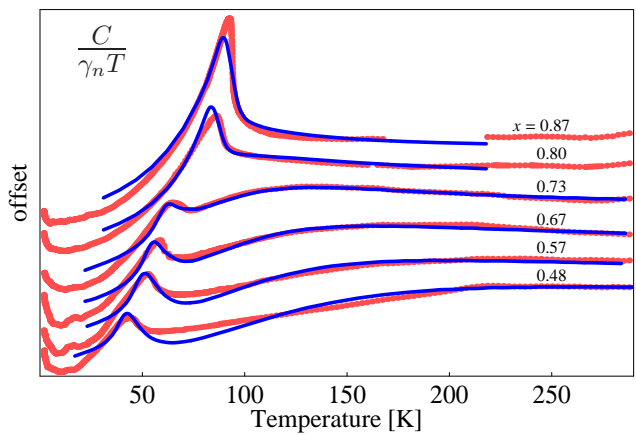}}}
 \put(1,4.5){\Large $ C \over \gamma_n T$}
 \end{picture}
\end{center}
\caption{YBa$_2$Cu$_3$O$_{6+x}$ specific heat  for different oxygen dopings $x$ compared to the average value method.}
\label{specificheat2}
\end{figure}
\index{YBCO}
\begin{figure}[ht]
\begin{center}
 \setlength\unitlength{1cm}
 \begin{picture}(8.5,5.5)
 \put(0,-0.5){\resizebox{8.5cm}{!}{\includegraphics{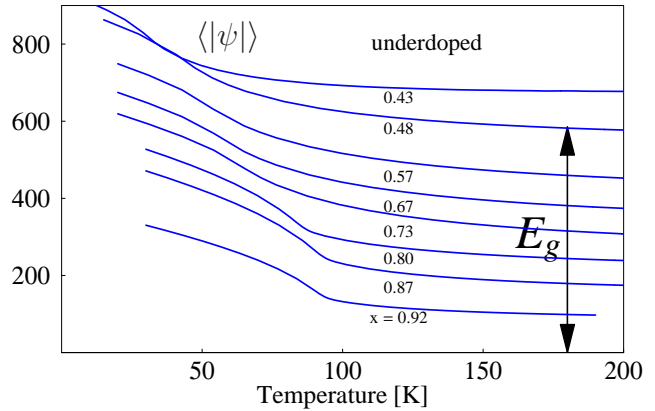}}}
 \put(2.5,4.5){\large $ \langle |\psi| \rangle$}
 \end{picture}
\end{center}
\caption{Average amplitude corresponding to YBa$_2$Cu$_3$O$_{6+x}$ specific heat of Fig. \protect{\ref{specificheat2}}
 for different oxygen dopings $x$. The amplitude for doping $x \approx 0.9$ resembles more to BCS behaviour whereas the
  amplitude for underdoped system is always very large due to amplitude fluctuations and large value of $T_0$. The pseudogap
  energy scale  $E_g$  is shown for doping $x=0.48$.
  }
\label{AbsoluteAmplitudeDoping}
\end{figure}

\begin{figure}[ht]
\begin{center}
 \setlength\unitlength{1cm}
 \begin{picture}(8.5,5.5)
 \put(0,-0.5){\resizebox{8.5cm}{!}{\includegraphics{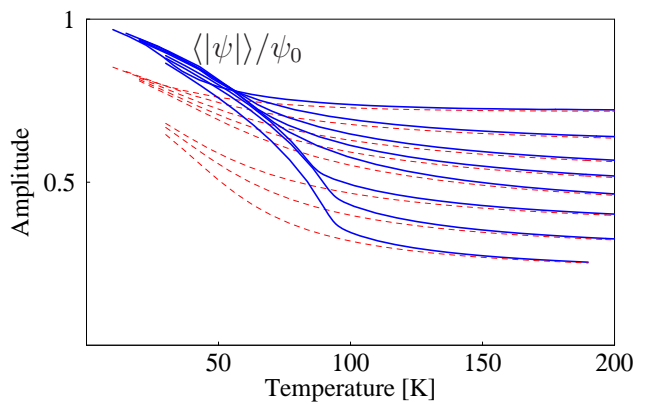}}}
 \put(2.5,4.1){\large $ \langle |\psi| \rangle/\psi_0$}
 \end{picture}
\end{center}
\caption{Reduced average amplitude corresponding to YBa$_2$Cu$_3$O$_{6+x}$ specific heat  for different oxygen dopings $x$.
 The dashed lines are average amplitudes calculated for a completely disordered phase field. $T_phi$ is the temperature where the average
 disordered amplitude deviates from the average amplitude, i.e. when effects from phases are visible. Dopings are the same as in figure
  \protect{\ref{AbsoluteAmplitudeDoping}}}.
\label{reducedAmplitudeDoping}
\end{figure}
\begin{table}
\begin{center}
\begin{tabular}{|p{2cm}|p{1cm}|p{1cm}|p{1cm}|p{1cm}|}
\hline
{\bf Doping }  & $T_c$[K]       &  $V_0$[K]    &  $T_0$[K]         &  $\alpha$       \\ [6pt]\hline
0.92               &  92.9           & 261.9          & 124.3               &  13.7                  \\ \hline
0.87               &  93.7           & 178.0          &  143.3              &   14.3         \\ \hline
0.80               &  88.8                & 121.8          & 158.9               &    10.6        \\ \hline
0.73               &  69.3           & 111.0          & 177.6               &   13.7              \\ \hline
0.67               &  60.8           & 88.9            & 193.0               &   18.1        \\ \hline
0.57               &  55.9           & 76.1            & 213.5               &    20.6         \\ \hline
0.48               &  47.7           & 57.6            & 241.5               &    16.0     \\ \hline
0.43               &  31.5           & 37.3            & 251.2               &    9.0     \\ \hline
\end{tabular}
\end{center}
\caption{ Extracted parameters $V_0, T_0, \alpha$ from the fits of YBCO specific heat. (see Figure \protect{\ref{specificheat2}}).}
\label {tableResults}
\end{table}
The average value method is compared to specific heat obtained for different doping in Fig. \ref{specificheat2}.
 For underdoped systems $x<0.80$ we use simulations in $d=2$. For the more overdoped, $x\geq 0.80$, simulations
  are done in $d=3$. The parameter $\eta$ is fixed to 3. Parameters $V_0$  and $T_0$ extracted from the fits are shown
   in the phase diagram of Fig. \ref{phasediagram2} and in table \ref{tableResults}. Values of the phase specific heat
    normalisation constant  $\alpha$ range from 10 to 20. The fitting procedure is done completely automatically by a
	 random walk in the parameter space  $\{V_0,T_0,\alpha\}$, until the error between experimental data and the fit is minimal.
	  As usual, a local minimum can be reached by using this procedure, and one can be trapped in this minimum. However,
	   when the dimension of the parameter space is 3 as in our case, different local minima can be easily excluded and the best fit can be achieved.

The temperature $T_\phi$ is derived from figure \ref{reducedAmplitudeDoping} where the average amplitude is compared to the
average amplitude $\langle | \psi| \rangle_r$ computed for a field where phases are random, i.e. the gradient of phases is set to $\pi/2$:
\beq
\langle |\psi| \rangle_r = {1 \over Z_r} \int d|\psi| \ |\psi| \ |\psi| \ e^{-{V_0 \over T}  \left[ \eta^2  ( t |\psi|^2 + {1 \over 2} |\psi|^4) + D |\psi|^2 \right] }
\label{randomamplitude}
\eeq
where $D$ is the dimension of the system and the normalisation factor $Z_r$ is  
\beq
Z_r =  \int d|\psi| \ |\psi|  \ e^{-{V_0 \over T}  \left[ \eta^2  ( t |\psi|^2 + {1 \over 2} |\psi|^4) + D |\psi|^2 \right] }
\eeq

\begin{figure}
\begin{center}
 \setlength\unitlength{1cm}
 \begin{picture}(8.5,5.5)
 \put(0,-0.5){\resizebox{8.5cm}{!}{\includegraphics{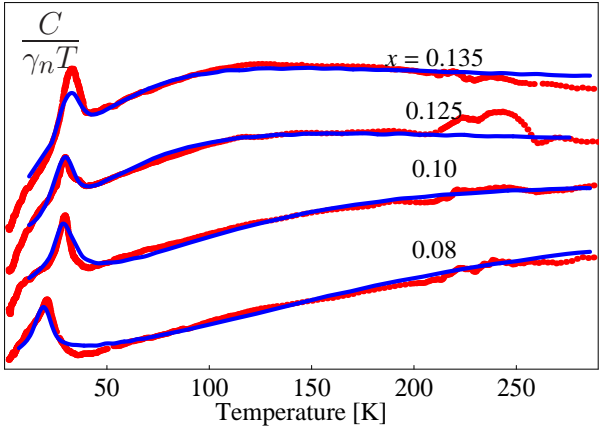}}}
 \put(0.5,4.5){\LARGE $ C \over \gamma_n T$}
 \end{picture}
\end{center}
\caption{Underdoped La$_{2-x}$Sr$_x$CuO$_4$ specific heat  for different strontium dopings $x$ compared to the $d$-wave average value method.}
\label{specificheatLSCO}
\end{figure}

\subsection{Specific heat for $s$-wave symmetry} 
\begin{figure}
\begin{center}
 \setlength\unitlength{1cm}
 \begin{picture}(8.5,5)
 \put(0,-0.5){\resizebox{8.5cm}{!}{\includegraphics{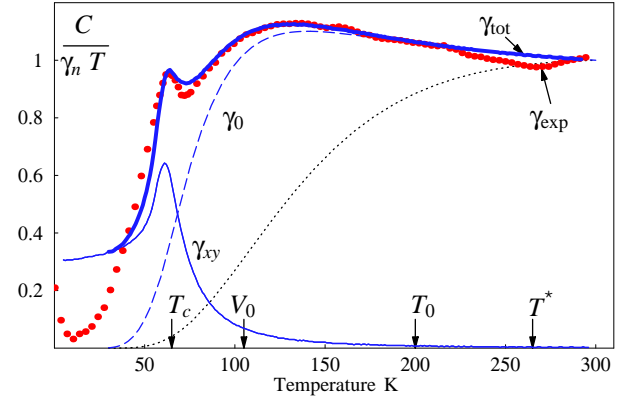}}}
 \end{picture}
\end{center}
\caption{Measurements (points) of the specific heat of YBa$_2$Cu$_3$O$_{6.76}$ divided by $\gamma_n
T $. The total $s$-wave specific heat (thick blue) is the sum of the critical XY contribution (thin blue) and
the amplitude contribution (dashed blue). If the amplitude of the gap were constant of size $2.14 T_0$,
the contribution would be the dotted black line.}
\label{specificheatswave}
\end{figure}

\index{s-wave}
The measured {\bf specific heat } of YBa$_2$Cu$_3$O$_{6.76}$ \cite{loram94} is compared to our
results for the $s$-wave symmetry in Fig. \ref{specificheatswave} by using the average value method.

%

\noindent
We took the following values for the two parameters: $T_0 = 235 K$ is of the order of $T^*$, $ V_0 =
108 K$ is of the order of $T_c$. For the size of $\psi$ we took the BCS $d$-wave relation:
$\psi(T=0)  = 2.14 T_0$ since it seems to be more in agreement with experiments  \cite{kugler} than the BCS relation of  $\psi(T=0)  = 1.76 T_0$.
This choice plays only a little quantitative role.

\begin{figure}
\begin{center}
 \setlength\unitlength{1cm}
 \begin{picture}(8.5,6.6)
 \put(0,-0.5){\resizebox{8.5cm}{!}{\includegraphics{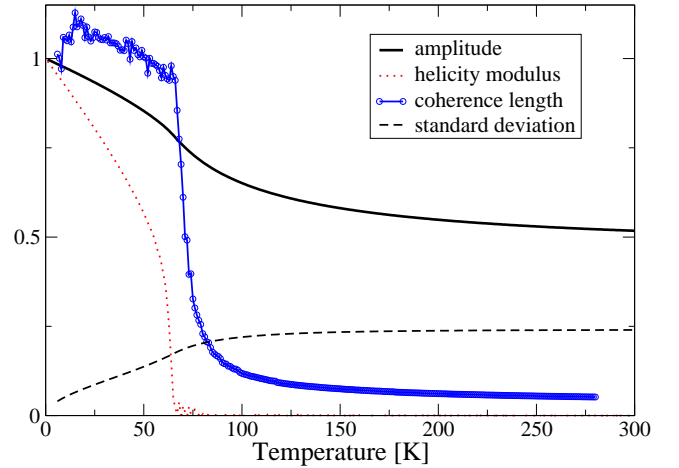}}}
 \end{picture}
\end{center}
\caption{ \label{amplitude} The average amplitude $\langle \R \rangle$ (thick) is large even above $T_c$ whereas the
coherence length $\xi$ vanishes rapidely above $T_c$. The standard deviation $(\langle \lt| \psi \rt|^2 \rangle - \langle \R \rangle^2)^{1/2}$
(dashed line) is large and almost constant. The phase stiffness $\Gamma_x$ in $x$ direction (dotted red) jumps from $2 / \pi$ to zero at $T_c$.
Lattice size: $N = 80^2$. Parameters are $T_0 = 235$ K, $V_0 =108$ K, $\eta^2 = 8$.}
\end{figure}

 The mean amplitude, standard deviation and phase stiffness are shown in Fig.
\ref{amplitude} for the same parameters as in Fig. \ref{specificheatswave}. The coherence length $\xi$ is of the order the lattice constant
when the temperature $T_\phi$ is reached.

%
Considering Fig. \ref{specificheatswave}, we see that an $s$-wave computation fits equally well experiments as the $d$-wave in
Fig. \ref{specificheat2}. Hence, we cannot decide which symmetry is favored by looking at specific heat data. $s$-wave and
$d$-wave symmetries are essentially different at low temperatures. $d$-wave specific heat has an algebraic increase with
temperature whereas $s$-wave specific heat has an exponential behaviour. Therefore the $s$-wave contribution is smaller
at low temperature than the $d$-wave. However, a lower $s$-wave contribution can be compensated by changing parameters $T_0$ and $V_0$.

\section{Differential conductance}
\index{differential conductance}
\index{tunneling}
 The differential conductance between a normal metal and a superconductor is directly related to the density of
states and the amplitude of the pairing field by the standard formula:
\beq
{dI \over dV} = - G_{nn} \sim { N_s(\xi) \over N(0)}
                              { \partial f(\xi + e V ) \over \partial( e V )}
\label{dIdV}
\eeq
where $G_{nn}$ is the differential conductance between two normal metals. The reduced energy is $\xi=\varepsilon-\mu$.
The $s$-wave density of states is according to BCS theory:
 \beq
N_s(\xi) = { | \xi | / \sqrt{ \xi^2 - \langle \D \rangle^2}}
\eeq
Of course, $N_s(\xi)=0$ if $|\xi|< \langle \D \rangle$.

\begin{figure}[h]
\centering
\resizebox{8.5cm}{!}{\includegraphics{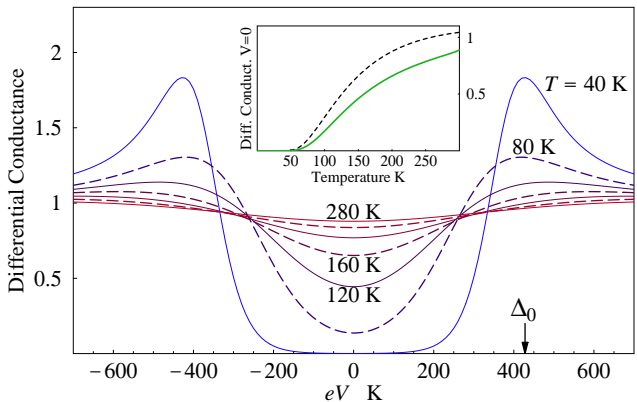}}
\caption{ \label{graphdIdV} The differential conductance of equation (\ref{dIdV}) shows a pseudogap, i.e a partial
suppression of the density of states, up to temperatures near $T^* \approx 300$. Inset: the differential conductance
(green curve) at $V =0$ recovers its normal behaviour near $T^*$. The low temperature approximation of equation
(\ref{lowdIdV}) is shown by the dashed line.}
\end{figure}

In Fig. \ref{graphdIdV}, the $s$-wave differential conductance normalised with $G_{nn}$ is presented using an average
 amplitude with the same parameters as in
Figure 4 of reference \cite{curty2003} ($T_0=159.2$ K, $V_0 =59.3$) for temperatures from 20 K to 300 K with 40 K intervals.
One can see that the width of the
differential conductance is proportionnal to the amplitude $\langle \D \rangle$, and that the gap fills up due to
thermal energy. Indeed, the low temperature differential conductance at $V=0$ is approximately given by:
\beq
{dI \over dV }(V=0) \approx G_{nn} \sqrt{2 \pi \langle \D \rangle \over T} \quad e^{-\langle \D \rangle /T} \label{lowdIdV}
\eeq
Theses results are in agreement with scanning tunneling microscopy on Bi$_2$Sr$_2 $CaCu$_2$O$_8$ of Renner {\em et al}
\cite{renner} where it is observed that the pseudogap gradually fills up whereas its width remains constant.

Do phase fluctuations contribute to the pseudogap? Yes but only for $T<T_\phi$ where the coherence length $\xi$ is of the
order of the lattice spacing as seen in Fig. \ref{amplitude}. As shown by Eckl {\em et al} \cite{eckl}, a constant amplitude of size $2 T_0$
maintains a pseudogap in the density of states up to $T^*$ whereas phase effects disappear near $T_c$.

\section{Extracted Phase diagram}
\begin{figure}[h]
\includegraphics[width=0.8\columnwidth]{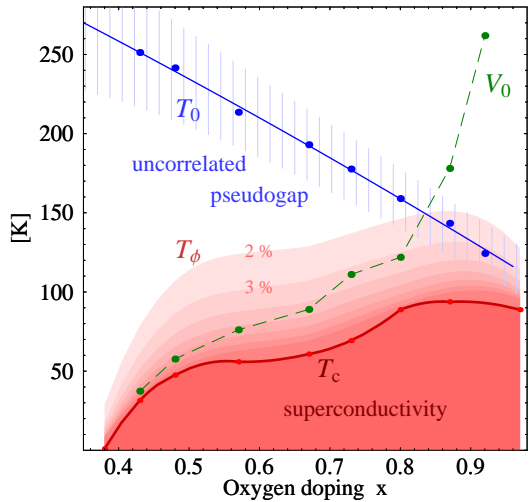}
\caption{Extracted phase diagram of YBa$_2$Cu$_3$O$_{6+x}$. Effects of amplitude  fluctuations are large in the quasi uncorrelated
 pseudogap region (below $T_0$) of the copper oxides phase diagram, whereas phase correlations remain important
  only below $T_{\phi}$.
 The temperature $T^*$,  where observables cross over to normal behaviour, is located in  the hatched area. }
\label{phasediagram2}
\end{figure}

Values for $T_0$, $V_0$ and $T_\phi$ extracted for the specific heat in Fig. \ref{specificheat2} are reported in
 Fig. {\ref{phasediagram2}}.
$T_\phi$ is obtained by using the comparison between amplitude from simulations and amplitude in
 the disordered phase as shown in Fig. \ref{reducedAmplitudeDoping}. $T_\phi$ is defined as the temperature where
  $\langle \D \rangle / \langle | \psi| \rangle_r = h $ for various thresholds $h =2\%, 3\%, ...$. The random amplitude
    $\langle | \psi| \rangle_r$ comes from equation (\ref{randomamplitude}).

 It is remarkable that phase correlations above $T_c$ grow rapidely in the underdoped regime following the
  $T_\phi$ line, and reduce when approaching the overdoped regime. $T_\phi$ lines are similar to Nernst effect
   results \cite{wang}. $T_\phi$ is also in agreement with the evolution of the temperature $T_{scf}$ (scf=superconducting fluctuations) of S. H. Naqib {\it et al} \cite{naqib} derived by   transport measurements.
   The gradient specific heat seems to disappears more rapidely. This latter doping dependence is
   in better agreement with the phase diagram derived in Hall effect experiments \cite{matthey}. \\

In fig. \ref{phasediagram2}, the pseudogap energy scale $E_g$ is reported in the phase diagram of YBCO.
  The pseudogap energy scale $E_g$ is defined here as the amplitude at $T=200\mbox{K}$:
\begin{equation}
 E_g =\langle |\psi|\rangle_{T=200\mbox{\footnotesize K}}
 \end{equation}
 $E_g$ shows approximately the same doping dependence as the one found by Loram and Tallon \cite{loram2}.
 However $E_g$ is not related to some additional hidden order parameter since $E_g$ is simply related to amplitude fluctuations as shown
 by calculations.

\begin{figure}
\includegraphics[width=0.8\columnwidth]{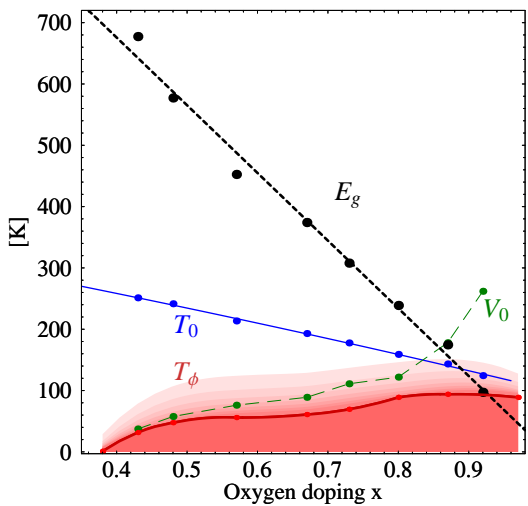}
\caption{The pseudogap energy scale $E_g$ in the extracted phase diagram of YBa$_2$Cu$_3$O$_{6+x}$.
}
\label{phasediagram3}
\end{figure}

\section{Discussion}
We have show that amplitude and phase fluctuations can explain the
emergence of the pseudogap region of underdoped high temperature superconductors. Phase
coherence disappears completely near a temperature $T_{\phi}$ above $T_c$, and therefore,
for $T > T_{\phi}$, the pseudogap region is dominated by amplitude fluctuations.
We find that the mean field temperature $T_0$ has a similar doping dependence as $T^*$, signaling that
 the pseudogap region is due to independent fluctuating pairs. The energy scale $E_g$ of the pseudogap
  is also derived. The large value of $E_g$ in the underdoped domain is due to a finite $T_0$ and large
  amplitude fluctuations. At maximum doping, $T_0$ remains of the order of $T_c$ but amplitude fluctuations are much
  weaker. Therefore $E_g$ vanishes linearly when approaching maximum doping. The fact that  $E_g$ takes small
  values above maximum doping should not be interpreted as a quantum critical point but as a crossover:
  amplitude fluctuations are still present in the
  overdoped regime and $E_g$ has always a finite value.\\

Comparison with measured specific heat on underdoped LSCO reproduces the double peak structure like in YBCO: a sharp
peak below $T_{\phi}$ coming from phase fluctuations and a separate wide hump
below $T^*$ rounded by the amplitude. The spin susceptibility, related to the
amplitude, recovers its normal behaviour near $T^*$ whereas the orbital magnetic
susceptibility, related to phases, disappears near $T_{\phi}$. These considerations
are independent of the underlying pairing mechanism, and any microscopic theory inducing
 pairing should lead to similar conclusions.\\

An important difference between overdoped and underdoped superconductors is the "separation" between effects of amplitudes and
 phases: in the overdoped regime, contributions of amplitude and phase are added producing only one peak
 in the specific heat for example. In the underdoped regime, phase correlations remain near $T_c$ and
 still produce a small peak whereas amplitude fluctuations extend to much larger temperature producing
  a separate wide hump between $T_c$ and $T^*$.\\

Further work is needed in order to include in the theory effects of the magnetic field on the pseudogap.

This work has been supported by the Swiss National Science Foundation.

\bibliographystyle{h-physrev4} 

\end{document}